# Atomistic mechanisms of binary alloy surface segregation from nanoseconds to seconds using accelerated dynamics

Richard B. Garza[1,2], Jiyoung Lee[3,4], Mai H. Nguyen[3], Andrew Garmon[5,6], Danny Perez[5], Meng Li[2], Judith C. Yang[2], Graeme Henkelman[3,4], Wissam A. Saidi[1*]

[1]Department of Mechanical Engineering and Materials Science, University of Pittsburgh, Pittsburgh, PA
[2]Department of Chemical and Petroleum Engineering, University of Pittsburgh, Pittsburgh, PA
[3]Department of Chemistry, University of Texas at Austin, Austin, TX
[4]Oden Institute for Computational Engineering & Sciences, University of Texas at Austin, Austin, TX
[5]Theoretical Division T-1, Los Alamos National Laboratory, Los Alamos, NM
[6]Department of Physics & Astronomy, Clemson University, Clemson SC

* Corresponding author. E-mail: alsaidi@pitt.edu

**Abstract**

Although the equilibrium composition of many alloy surfaces is well understood, the rate of transient surface segregation during annealing is not known, despite its crucial effect on alloy corrosion and catalytic reactions occurring on overlapping timescales. In this work, CuNi bimetallic alloys representing (100) surface facets are annealed in vacuum using atomistic simulations to observe the effect of vacancy diffusion on surface separation. We employ multi-timescale methods to sample the early transient, intermediate, and equilibrium states of slab surfaces during the separation process, including standard MD as well as three methods to perform atomistic, long-time dynamics: parallel trajectory splicing (ParSplice), adaptive kinetic Monte Carlo (AKMC), and kinetic Monte Carlo (KMC). From nanosecond (ns) to second timescales, our multiscale computational methodology can observe rare stochastic events not typically seen with standard MD, closing the gap between computational and experimental timescales for surface segregation. Rapid diffusion of a vacancy to the slab is resolved by all four methods in tens of ns. Stochastic re-entry of vacancies into the subsurface, however, is only seen on the microsecond timescale in the two KMC methods. Kinetic vacancy trapping on the surface and its effect on the segregation rate are discussed. The equilibrium composition profile of CuNi after segregation during annealing is estimated to occur on a timescale of seconds as determined by KMC, a result directly comparable to nanoscale experiments.

1. **Introduction**

Alloy surfaces are typically enriched with one of their constituent elements, particularly in the top layers, due to differences in the surface energies of the pure metals. This surface segregation process leads to metallic de-mixing, which is of relevance to many different fields of research such as catalysis and metallurgy, considering that in situ transformations can affect the chemical activity or structural integrity gained from homogeneously alloying pure metals together. Alloy surface segregation and ordering have been measured experimentally both in vacuum and gas environments, with differences in the equilibrium alloy composition induced near the exposed top layers for many bimetallic alloys[1-6]. Consequently, studies of alloy surface transformations are influenced by prior phase separation in vacuum during pre-treatment. A decoupling of the experimental environment from the pre-treatment environment is required to understand the transient effect on non-equilibrium surface composition, since nanoscale elemental mapping is not feasible on relatively short microsecond (μs) timescales even using electron and X-ray diffraction.

Molecular dynamics (MD) simulations can generally resolve atomic transitions on the picosecond (ps) to nanosecond (ns) timescales. While these short timescale dynamics can be easily investigated using energetic models based on either classical force fields or first-principles methods, longer timescales that are often more relevant experimentally are challenging to realize with conventional resources and techniques[7-9]. As shown in the current work, alloy surface segregation occurs over millisecond (ms) timescales that are impractical to obtain using conventional MD simulations. This goal can be achieved with accelerated methods including adaptive kinetic Monte Carlo (AKMC) and temperature-accelerated dynamics (TAD)[10-12]. These two methods were employed in a multiscale approach to study surface segregation in ceramics (rocksalt oxides) during oxidation[13]. Also, more recently, surface segregation and timescales in PdAu nanoparticles have been studied using AKMC[14]. However, alloy segregation and timescales related to the metallic dopant dynamics are not well understood in nanoscale thin films, as these require larger model systems to accurately simulate the region between bulk and exposed surface. A recent KMC study investigated the dynamics in bulk NiFe solid solution, reporting that the vacancy migration barrier is highly influenced by the local composition, behavior that the authors predict will influence the rate of microscale phase transformations[15]. Previously, only the dependence of activation energy on overall alloy composition could be predicted [16]. The failure of these global models for concentrated solid solutions is known, though they may be corrected by sampling transition energies in differing chemical environments[17].

In this study, we model the non-equilibrium surface segregation process by employing multiscale simulations methods that probe the dynamics from ps to second timescales. We focus our investigations on cupronickel (CuNi), which exhibits surface segregation at elevated temperatures in vacuum, enriching with Cu near the surface and Ni in the bulk[18, 19]. The CuNi alloy is of interest for different applications with extreme environmental conditions including marine settings due to its resistance to corrosion by seawater[20, 21], as well as a high-temperature catalyst for thermal $CO_2$-to-syngas-to-fuels conversion[22-25].

To date, the transient dynamics of segregation in CuNi have not been investigated. Previous studies have employed Monte Carlo (MC) to show that the Ni solute concentration is significantly decreased in the top three surface monolayers for all slab orientations, approaching its value in the bulk at a depth of four or five atomic layers below the surface[26-28]. While MC simulations describe the systems in equilibrium, they do not give a timescale for the segregation process. In contrast, MD can provide a timescale for dynamics, but it has not been utilized before to study transient dynamics of the segregation process. Limited investigations of dislocation slip under applied stress (work hardening) or melting in CuNi [29, 30] have been carried out using MD because these processes occur over ns timescales, yet the impact on larger scale reordering on experimental timescales was not explored in those studies. Thus, the mechanism, as well as the timescale, for transformation from a randomly mixed to ordered alloy on the nano- and microscales remains unclear.

Accelerated methods can provide insight on longer timescales into atomistic mechanisms of surface segregation phenomena. Previously, AKMC investigations of the segregation kinetics of PdAu nanoparticles showed greater kinetic stability due to reduced strain in the mixed-phase[14]. In the present work, we probe segregation dynamics in planar, (100) CuNi surfaces, employing conventional MD and three accelerated dynamics methods: parallel trajectory splicing (ParSplice), AKMC, and KMC with kinetic barriers derived from a cluster expansion. ParSplice affords accurate system evolution up to ~10 μs, while AKMC and KMC simulate longer timescales up to ms and seconds, respectively.

ParSplice [11] extends MD simulation times by leveraging parallel computers to carry out parallelization in the time domain, in contrast to the usual domain decomposition approaches that operate in the space domain [31, 32]. Thus, with ParSplice it is possible to simulate small systems over very long timescales, again in contrast to conventional parallelization approaches that are efficient at spatially decomposing large systems simulated over short timescales. This is accomplished by concurrently generating a large number of independent, short trajectory segments using a procedure that guarantees these segments can be assembled into a longer state-to-state trajectory that is statistically correct[33]. Periodic quenching is used to identify transitions between different metastable states and segment terminations. It can be shown that ParSplice trajectories can become arbitrarily accurate by adjusting the estimate of the so-called correlation time of the dynamics, at the expense of a computational overhead [34]. When the dynamics follow from a sequence of rare events, ParSplice can provide a computational speedup that scales with the number of processors used; it is therefore especially powerful when deployed on massively parallel computers.

KMC-based methods do not have a fixed timestep; instead, they find the time elapsed for the first escape from one state to another, allowing for large periods of vibrational motion in the atomic system to be bypassed[14, 35]. These escape times correspond to reaction rates, which are calculated adaptively or "on the fly" to construct an AKMC state model: nothing in the output event table is predefined or assumed from prior knowledge[12, 14, 35]. AKMC uses minimum-mode following searches or high-temperature MD to construct this event table: the transition state energies (activation barrier heights) are found using single-ended saddle point finding algorithms such as the dimer method[36-38]. To achieve further acceleration while maintaining as much of this accuracy as possible, in the present study, we use off-lattice dynamics from AKMC to fit a more approximate, lattice-based KMC model, which functionally depends on the Ni-coordination via a cluster expansion.

Our multiscale approach produces a hierarchy of trajectory data for the segregation process over a broad range of timescales. To represent trajectory data on varying scales and fidelities, we track physical properties including the local defect chemical environment and its influence on the correlated rates of segregation and vacancy migration. Further, we find that segregation in the top layer is a function of the rare re-entry of a vacancy from the surface to the subsurface, which only occurs with a frequency of 10 µs due to a high kinetic barrier, greatly increasing the time required for the surface to be completely depleted of Ni. We also determine that the number of dopant atoms in the $1^{st}$ and $2^{nd}$ coordination shells around the point defect alters its migration energy, affecting the rate of composition change. The consistency of system evolution during segregation by thermal annealing across all the accelerated methods is examined on many timescales; this evidence supports each technique's further use in multiscale simulations in combination with the data processing methods used in this work.

## 2. Methodology

*2.1 Validation of the Embedded-Atom Method Potential*

A reliable interatomic potential is required to obtain accurate MD of surface segregation and equilibrated structures. Here, we employed the embedded atom model (EAM) potential of Fischer and collaborators, which is designed to model CuNi phase segregation across grain boundaries and is parameterized with surface energies, lattice constants, vacancy migration energies, and relevant quantities governing the rate of metallic phase separation [39]. We first verified the applicability of this potential to study CuNi surface segregation by comparison of calculated surface energies with those obtained from spin-polarized DFT calculations carried out using VASP with the Perdew-Burke-Enzerhof (PBE) exchange-correlation functional[40-44], see SI Appendix A, Table A1. Further, we verified that the segregation of Ni to the bulk—from the 1st to the 2nd or 3rd layer—is always favorable i.e. $\Delta E_{Ni,surf \rightarrow interior} \approx$ -0.4 to -0.3 eV from DFT and EAM, also in agreement with previous ab initio calculations [45]. Further, we found good agreement with the previously reported anisotropic, thermodynamic tendency for Ni to segregate to exposed <100> and <110> facets in favor of <111> surfaces[19, 27].

*2.2 Slab Models and Initial State for Annealing*

We employed slab models of the (100) surface termination that are composed of 216 and 384 atoms by randomly substituting Cu with Ni. These slabs were 12 monolayers in thickness with 3x3 and 4x4 surface periodicity; nearly doubling the vacancy concentration (1:216 vs. 1:384 atoms) did not affect our observed mechanisms or energetics. Thus, systems of smaller surface periodicity (3x3, 216 atoms) were simulated to realize the longest timescales for higher fidelity time-averaging and activation energy histograms. MC simulations across the composition range of 2.7 – 16at% Ni, and temperature range of 300 – 700 K showed no detectable variation in the amount of Ni segregated. For this reason, we chose to exclusively simulate systems with 16at% Ni at 500 K with the accelerated MD methods.

The equilibrium composition profile of the CuNi (100) surface alloy is determined using MC. The method attempts $10^6$ MC swaps of Cu and Ni atoms with subsequent minimization and an acceptance probability as determined via the Metropolis algorithm[46]. MC composition profiles are found to be consistent with similar reported profiles [18, 26-29, 47] across a range of compositions and temperatures, see SI Appendix B. Such agreement with previous computational and experimental results is a further validation of the reasonable accuracy of the EAM potential.

*2.3 Multiscale Simulation Hierarchy*

The methods employed in our multiscale hierarchy are described as follows, in order of accessible timescale: molecular dynamics in the NVT ensemble was generated using a Langevin thermostat, as implemented in the LAMMPS code[48]. Many instances of LAMMPS were then orchestrated by the EXAALT/ParSplice code [http://gitlab.com/exaalt] to carry out long-time ParSplice simulations. Direct MD realized ns of simulated time using the velocity Verlet algorithm and modest resources (4 processors). ParSplice accelerated these dynamics up to μs timescales using 224 processors. Both standard MD and ParSplice were carried out in the same way: after initializing particle velocities according to a Boltzmann distribution representing the target temperature, the systems were equilibrated with a 2 fs timestep and a 1 ps temperature relaxation time for 200 ps. After the equilibration phase of MD, a further 400 ns were simulated to constitute the production phase, while ParSplice accelerated this simulation time up to 35 μs. Minimization was done with the conjugate gradient algorithm using convergence criteria of $10^{-8}$ eV and $10^{-6}$ eV/Å for the energy and forces[49], respectively. Quenching is necessary for ParSplice to identify state transitions and to properly terminate each parallel MD replica.

In contrast to ParSplice and MD, the AKMC and KMC models use a non-constant timestep, which varies to match the timescale of the first escape time from a given state. Over longer timescales, up to many seconds, KMC-based acceleration approaches the segregation found with MC [12, 35]. KMC methods simulate the time evolution of the system, requiring a predetermined event table in which the kinetic rate of each event is approximated through the Arrhenius relation to the pre-calculated activation energy. At each step, a random event $i$ is selected from the table in the order from 1 to $i$ with the condition

$$\sum_1^{i-1} r_{i-1} < p_1 \sum_1^N r_N \leq \sum_1^i r_i$$

where $\sum_1^i r_i$ is the sum of the rate from event 1 to event $i$, $\sum_1^{i-1} r_{i-1}$ is the sum of the rate from event 1 to event $i$–1, $p_1$ is a random number between 0 and 1, and $\sum_1^N r_N$ is the total rate of the event table. Time is then incremented by

$$\frac{-\ln(p_2)}{\sum_1^N r_N}$$

where a random number $p_2$ is drawn between 0 and 1. AKMC allows the system to find all potential events without a need of a predetermined event table [45]. In order to search for events, AKMC uses high-temperature MD and the climbing-image nudged elastic band (CI-NEB) approach to calculate the saddles for the new states using the EON software[50].

To reach even longer timescales with KMC, an energy estimation based on the local environment of the vacancy was generated by a cluster expansion. The energy was predicted to be dependent on the concentration of Ni and Cu located in the first nearest neighbor shell of the vacancy. In the FCC CuNi alloy, there are eight nearest neighbors for the vacancy on the surface and twelve nearest neighbors for the vacancy in the subsurface. Sequentially, the energy-fitting model was used to determine the barrier and rate of the events in the event table based upon the trajectory data from our AKMC simulations. More information about the cluster expansion method is provided in the SI, Appendix C. Further, we found satisfactory agreement between AKMC and KMC by carrying out the dynamics up to 300 μs, further validating the KMC model (see SI, Appendix D).

Finally, the concept of an equilibrium rate was introduced to further accelerate the timescale accessible by KMC. In this approximation, no rate was allowed to be larger than the specified equilibrium rate with the assumption that all states connected by rates faster than the equilibrium rate should already be in equilibrium. In the case of CuNi segregation, the planar diffusion of a vacancy on the exposed surface is rapid, equilibrating on much shorter timescales than for defect re-entry to the subsurface to occur even once. Ultimately, the effect of an artificial equilibrium rate in KMC simulations is that these key transitions can be sampled more effectively instead of the many horizontal transitions that do not alter the composition with respect to surface depth.

In both AKMC and KMC simulations, the temperature was set to 500K and the prefactor for the rates was fixed at $5\times10^{12}$ $s^{-1}$. The optimizer used in AKMC was L-BFGS, with a convergence criterion of 0.01 eV/Å[49]. System evolution from AKMC reached ms timescales running for a week on 24 cores. The following KMC timings are the average of five separate runs in both cases: KMC realized seconds of simulation time running on a single processor for 2 hours with the added equilibration rate. KMC without the equilibrium rate was only simulated up to 10 ms, since only one minute of wall clock time was necessary before the surface vacancy trapping described above halted Ni segregation in the system.

### 3. Results and Discussion

The equilibrium composition profiles obtained with MC (Figure 1) exhibit Ni migration out of the top three surface layers, while the concentration of the fourth and fifth layers approach the bulk value. This trend did not quantifiably vary with changes in system temperature or Ni concentration. The near-surface Ni concentration observed with the EAM potential agrees with the profiles derived from previous MC simulations [18, 26-29] as well as experiments[1-3, 19].

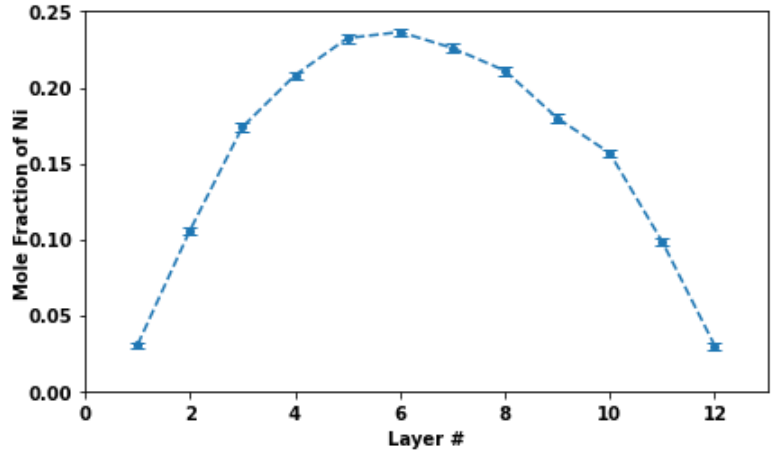

**Figure 1:** Ni composition profiles of Cu-16at%Ni (100) surface as a function of layer depth after MC annealing at 500 K. The slab has 12 layers with layers 1 and 12 exposed to the vacuum. Composition is normalized by the number of atoms in a pristine FCC <100> layer (18) resulting in a mismatch between the bulk composition of layer 6 (~22at% Ni) and the overall composition (16at% Ni).

Next, we probed the segregation dynamics using MD, ParSplice, AKMC, and KMC starting from the same initial configuration of the random alloy. Lattice vacancies are the primary defect responsible for alloy segregation: the only other mechanism, self-interstitial migration of metal atoms to octahedral or tetrahedral sites, is destabilized by greater formation and migration energies than those for the point defect (vacancy)[51]. Hence, our simulations included a single vacancy to facilitate surface segregation.

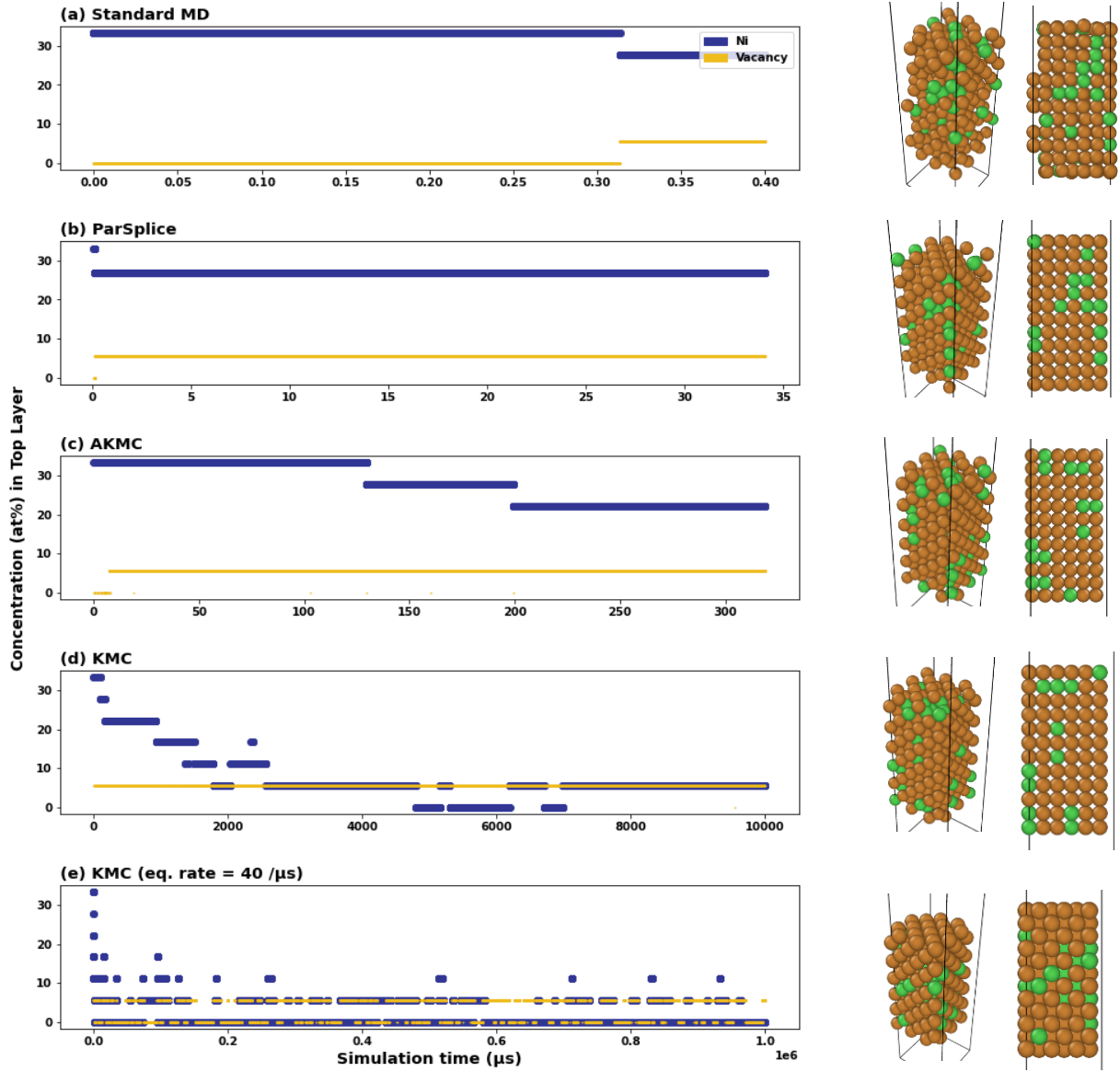

**Figure 2:** (left) Ni content in the top layer of Cu-16at%Ni(100) with a vacancy for (a) standard MD, (b) ParSplice, (c) AKMC, (d) KMC, and (e) KMC with applied equilibrium rate. Composition is normalized as in Figure 1. (right) Orthographic and side views of the final slab model from each simulation are provided for reach simulation.

Figure 2 shows the composition over time within the top layer of our Cu-16at%Ni (100) slab model for a single simulation of each type within our methodology. The concentration of Ni atoms in the top surface layer decreases over time as it transitions from a uniform distribution to that of equilibrium, as shown by the MC simulations in Figure 1. The accessible simulation time increases sequentially for each method (MD, ParSplice, AKMC and KMC). In MD (Figure 2a), the vacancy diffuses to the surface at 0.3 μs, displacing a Ni atom to the subsurface; up to the total time of 0.4 μs the vacancy remains trapped on the surface so that the Ni concentration does not change. With over thirty different MD simulations, we observed that the time for vacancy percolation to the surface was consistently less than 1 μs, resulting in similar degrees of segregation as shown in Figure 2a.

ParSplice (Figure 2b) shows similar dynamics as in standard MD, with vacancy migration to the surface in a fraction of a μs. Throughout the simulation, the ParSplice trajectory visited 8,278 topologically unique states while making 37,458 transitions, the vast majority of which occurred after the vacancy had reached the surface. The parallel acceleration of ParSplice increased the timescale from the standard MD simulation by a factor of ~100 up to 35 μs, but even at this longer timescale the vacancy remained trapped on the surface and no additional Ni segregation was observed. As expected, ParSplice is consistent with MD on timescales where they overlap (0.4 μs) not only in terms of bulk vacancy diffusion but also the timescale at which the vacancy diffuses to the surface. We have repeated the ParSplice simulations dozens of times with different random seeds observing early vacancy diffusion to the surface (before 1 μs) in nearly all trials, just as with MD.

With AKMC, we increased the simulation timescale by another order of magnitude (Figure 2c): a total of 144,597 transitions evolved the system through 32,369 unique states (a similar ratio of transitions to new states found as in ParSplice). Note that our AKMC approach uses coarse-graining following the MC with absorbing Markov chains (MCACM) method, allowing many more transitions to be considered via an analytic solution to the rate equations[52]. From the AKMC dynamics, we can observe events in which the vacancy moves from the surface to the subsurface at a timescale of roughly 50 μs. Over the simulation time of 300 μs, five such events were observed resulting in two Ni atoms and 3 Cu atoms migrating from the surface to the subsurface.

With our KMC model (Figure 2d), we executed 4.7 million transitions generating 10 ms of simulated time. Over these timescales, the system appears to approach equilibrium, with fluctuations in the surface concentration between 0 and 5% Ni. However, what is not obvious from these plots is that the vacancy spends all of its time in the first and second layer, so that Ni segregation only occurs between the surface and subsurface layers. This behavior originates from a disparity in barrier heights that embodies the "low-barrier problem", more aptly referred to as the "heterogeneous barrier" problem. Dynamics with a mix of low and high activation barriers are inherently more difficult to accelerate since groups of states interconnected by low barriers will always dominate the trajectory during naïve state space exploration[8, 11]. The barrier for vacancy surface diffusion is 0.4-0.5 eV, whereas the barrier for the vacancy to go subsurface is 0.8-0.9 eV. Thus, we are simulating on the order of a million KMC steps with the vacancy mostly diffusing on the surface for one subsurface diffusion event, offering a low chance for segregation to occur. Even factoring in the small cost of each KMC step, this makes it impossible to simulate an equilibrium distribution of Ni in the top three layers.

To further accelerate the dynamics and mitigate the "heterogeneous barrier" problem, in our final simulation (Figure 2e), we perform KMC with the equilibrium rate approximation described in the methods section, realizing 160 million transitions to reach a simulated time of 1 s. Here, the idea is that the vacancy will quickly reach local equilibrium diffusing in the top layer, and no new states of interest are explored until subsurface diffusion occurs. Since vacancy surface diffusion occurs on a timescale of ns, and diffusion of the vacancy to the subsurface occurs on a timescale of μs, we chose an equilibrium rate of 40/μs to slow the surface diffusion and accelerate the diffusion of the vacancy to the subsurface and below. Figure 2e shows that on a timescale of seconds, the surface Ni concentration fluctuates around equilibrium after 0.4 seconds until the end of our simulation lasting 1 s.

In order to more accurately estimate the timescale required to obtain the equilibrium profile, we performed five individual KMC simulations with the same equilibrium rate (40/μs), extending the total time to 2.5 second corresponding to about 400 million transitions. Figure 3 shows the averaged Ni composition obtained from these KMC simulations in the bottom three layers (referred as layer 1, 2 and 3) and the top three layers (referred as layer 12, 11, and 10) with the equilibrium composition profiles obtained by MC together for comparison. From Figure 3a, we can see that layer 1 reaches the equilibrium profile in less than 0.1 second and even approaches zero Ni concentration at ~1.5 second. Layer 2 exhibits more dramatic changes. The composition started from 30% Ni, dropped to the equilibrium composition of 10at% ~1.75 second, and remained near the equilibrium until the end of the simulation. While the MC simulation left 16% Ni in layer 3, our KMC model shows more Ni segregations leaving 10% Ni in the layer. This might explain the higher Ni composition profiles of the top two sublayers (layer 11 and 10) in Figure 3b. While layer 12 shows similar behavior as layer 1, layer 11 did not reach equilibrium until 2.5 second, and layer 10 did reach the equilibrium at ~0.6 s before the composition increased again. The overall profile, nonetheless, clearly shows an increase in the number of Ni atoms in bulk layers, indicating thermodynamic tendency for this dopant to remain in the bulk rather than on the surface.

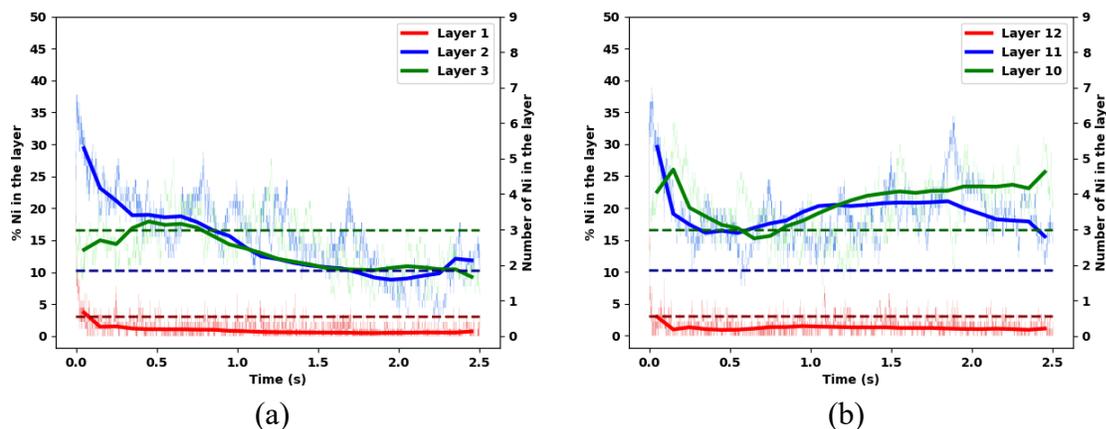

(a)                                    (b)

**Figure 3:** Ni composition averaged over five KMC simulations for the (a) bottom three layers and (b) top three layers over 2.5 second with an equilibrium rate of 40/μs. The bold line represents the 0.5-second-average calculated from the lighter single-frame datapoints, and the dashed line corresponds to the average concentration of Ni by MC calculation with EAM potential at each layer when the system reaches equilibrium. Composition is normalized as in Figure 1.

Apart from surface effects, the local chemical environment around a vacancy is expected to determine the system energetics and the rate of Ni composition change. To investigate this, we tracked the number of Ni atoms within a 5 Å radius of the vacancy. The number of Ni atoms in the local environment ranges from 0 to 12, with three distinct "bins" formed for low, mixed, and high Ni-content environments. The spectrum of barriers calculated during AKMC for all vacancy migration events within this trajectory is presented as a histogram in Figure 4a, which shows the lowest transition state energies for vacancy migration in Ni-rich regions of the alloy. Additionally, because the system has lower Ni content than Cu, the integrated peak area is smaller for transitions into/within Ni-rich regions than for those with mixed and Cu-rich compositions. Vacancy migration energies are shifted closer to 0.4 eV in the Ni-rich regions than for migration in Cu-rich regions according to Figure 4a. We can deduce that vacancy migration is favored in the Ni-rich regions, contributing to the lower dwell time near Ni as the vacancy more rapidly diffuses towards and away from this dopant. Vacancies must slowly explore the Cu-rich regions of the host lattice before returning to possibly segregate the Ni atoms away from the surface. The order of integrated peak areas in Figure 4a also supports this conclusion.

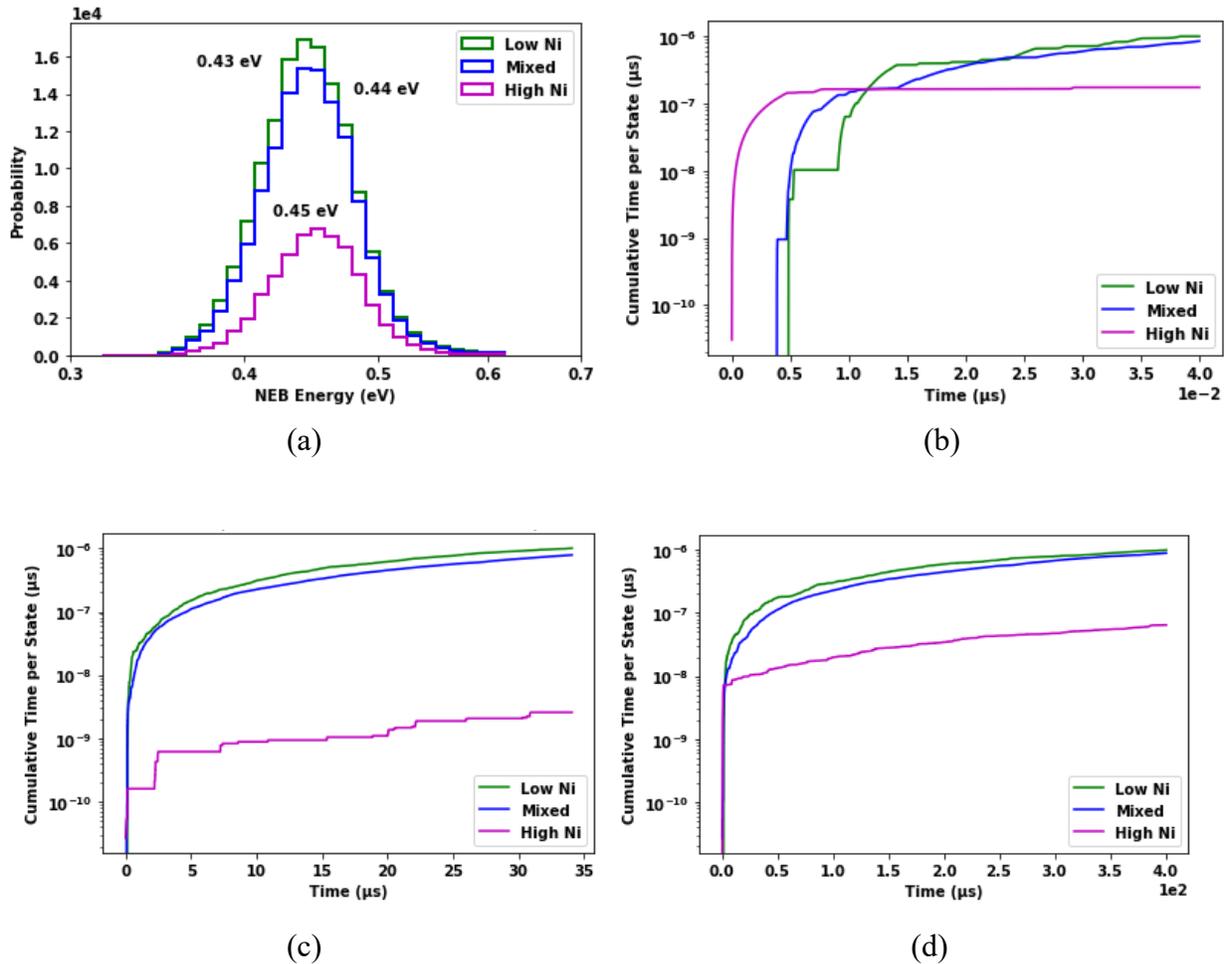

(a)  (b)

(c)  (d)

**Figure 4:** (a) Histogram of vacancy migration energy barriers obtained from a 319 μs AKMC simulation. (b-d) Total residence time for the vacancy in different chemical environments for (b) MD, (c) ParSplice, and (d) AKMC. Cumulative times are normalized by the prevalence of each composition type in the system rendering them unphysical.

The overall time spent in low Ni, mixed, and high Ni content environments is also shown with respect to time in Figure 4b-d using a cumulative measure of residence time for MD, ParSplice, and AKMC and re-scaling this measure by the prevalence of each environment in the system. The number of Ni atoms in the vicinity of the vacancy modulates these timescales substantially but in similar fashion for all methods. Further, the logarithmic trendlines and their orders of magnitude agree across overlapping timescales for MD/ParSplice (ns) and ParSplice/AKMC (µs): this supports the hierarchy of methods utilized herein as a theoretical foundation to connect correlated observations evolving across many timescales for the same process. The cumulative dwell time and thermodynamic trends match for all of the simulation methods in the order of integrated peak areas (Figure 4a) and residence times (Figure 4b-d): $t^{total}_{Cu-rich} > t^{total}_{mixed} > t^{total}_{Ni-rich}$. Though this agreement does not hold at very early (ps) simulation times with low cumulative sums, particularly for MD which is noisier as compared to accelerated methods, the trend becomes evident in the ergodic limit. The residence times of vacancy chemical environment presented in Figure 4b-d confirm that the local composition is an effective determinant of where a vacancy spends most of its time during annealing and segregation. Specifically, the position of Ni in the lattice is a minor determinant of the dynamics, since the dopant slightly biases the vacancy's random walk by ~0.1 eV. This effect has also been documented in Ni-Fe surfaces annealed at 1100 K using KMC, where the local composition and the identity of the atoms exchanging during segregation (solute vs. solvent) significantly influenced the vacancy migration energy and the measured tracer diffusion coefficient[15].

## 4. Conclusions

In summary, we have applied three accelerated dynamics methods to examine the rate of segregation in CuNi alloy from nanosecond to second timescales, reaching 1 second of simulation time and the equilibrium composition with KMC. This composition profile shows no Ni on the top surface monolayer, with less than 15at% Ni in the $2^{nd}$ layer compared to 22at% Ni in the bulk on a per layer basis, in agreement with MC predictions. Though most of our accelerated methods were used to simulate up to the µs timescale, only modified KMC dynamics could reach the equilibrium profile obtained from MC and previous experimental observations. Our model estimates that the timescale for segregation in the top layer to reach equilibrium is on the order of 0.1 ms, while equilibrium segregation does not penetrate to the $3^{rd}$ layer until timescales on the order of 100 ms. The equilibrium timescales for surface segregation of any FCC bimetal can be determined with a combination of AKMC and KMC. However, KMC-based methods require assumptions regarding transition state theory, and they do not resolve fast atomic exchange processes as well as ParSplice and MD. Our study shows that a model for the relationship between solvent distribution, activation energy, and large-scale phenomena like segregation may be developed from further simulations of bimetallics on experimental timescales.

## 5. Declaration of Interests

The authors declare that they have no known competing financial interests or personal relationships that could have appeared or considered to influence the work reported in this paper.

## 6. Acknowledgments

This work was funded from NSF (DMR-1809085, CHE-210231, and CMMI-1905647) as well as resources and collaboration from the University of Pittsburgh's Computational Resource Cluster

# Supplementary Information

# Atomistic mechanisms of binary alloy surface segregation from nanoseconds to seconds using accelerated dynamics

Richard B. Garza[1,2], Jiyoung Lee[3,4], Mai H. Nguyen[3], Andrew Garmon[5,6], Danny Perez[5], Meng Li[2], Judith C. Yang[2], Graeme Henkelman[3,4], Wissam A. Saidi[1*]

[1]Department of Mechanical Engineering and Materials Science, University of Pittsburgh, Pittsburgh, PA
[2]Department of Chemical and Petroleum Engineering, University of Pittsburgh, Pittsburgh, PA
[3]Department of Chemistry, University of Texas at Austin, Austin, TX
[4]Oden Institute for Computational Engineering & Sciences, University of Texas at Austin, Austin, TX
[5]Theoretical Division T-1, Los Alamos National Laboratory, Los Alamos, NM
[6]Department of Physics & Astronomy, Clemson University, Clemson SC

[*] Corresponding author. E-mail: alsaidi@pitt.edu

**Appendix A: EAM Validation Using DFT**

The surface energies of doped Cu surfaces up to 2.7at%Ni were computed using DFT and the

EAM potential of focus in this work: this was done not only to validate the forcefield's application in the study of surface segregation, but also to examine the process with ab initio accuracy[39]. We used the Vienna Ab Initio Package (VASP) [40-42, 44] in conjunction with Perdew-Burke-Ernzerhof (PBE) [43] exchange-correlation functional and an energy cutoff of 500 eV to expand the wavefunction. We used a 3 x 3 x 1 k-point mesh generated with a Monkhorst-Pack routine to sample the Brillouin zone for <100>, <110>, and <111> slab models composed of ~100 atoms. All calculations are spin polarized. We used a slab approach to model the surfaces and included 10 Å of vacuum between exposed surfaces to mitigate spurious interactions in the non-periodic direction. One of the host lattice atoms in the system (Cu) were replaced with Ni at both surface (1st layer) and subsurface (2nd layer) positions. The surface energy of a pure Cu slab model with (hkl) orientation including $N_{atoms}$ is calculated relative to the bulk energy per atom $E_{Cu}$ and the total exposed area 2A:

$$E_{hkl}^{surf} = \frac{E_f^{pure} - N_{atoms} E_{Cu}^{bulk}}{2A}.$$

The equation for surface energy of models containing one dopant atom is very similar to the above, and requires the bulk energy per atom of the dopant (Ni in this study):

$$E_{hkl}^{surf} = \frac{E_f^{doped} - (N_{atoms} - 1) E_{Cu}^{bulk} - E_{Ni}^{bulk}}{2A}$$

The pure Cu surface energies for orientations with low Miller index ($h^2+k^2+l^2 < 3$) are shown in Table A1, which were calculated with DFT and EAM. As is known from previous calculations and measurements of surface energy for pure Cu, <111> surfaces are most thermodynamically stable, followed by <100> and <110> surfaces. The validation with pure Cu surfaces finds 2.6% average relative error between the EAM potential and first-principles. We then replaced one atom in these structures with Ni to study the effect of doping at different surface depths.

|  | DFT (J/m$^2$) | EAM (J/m$^2$) |
|---|---|---|
| **Cu (100)** | 1.412 | 1.357 |
| **Cu (110)** | 1.477 | 1.488 |
| **Cu (111)** | 1.208 | 1.246 |

**Table A1:** Surface energies of pure Cu from DFT and the EAM potential of Fischer et al

Oriented slabs along (100), (110), and (111) planes exhibit segregation trends as a function of crystallographic direction in the FCC system for Ni, evident in Table A2. EAM calculations of Cu surface energy with an included Ni dopant at various locations have only 4.7% average relative error compared to DFT results. Additionally, the difference in minimized energies between structures with Ni on the exposed surface or beneath it (subsurface) is 0.38-0.39 eV for <100> and <111> Cu surfaces, yet this gap is only 0.17 eV for doped <111> Cu surface. The same trend was found to cause the equilibrium segregation profiles from previous MC studies of CuNi surface separation, agreeing with the calculated energetics in our own validation step[18, 26, 27, 53, 54]. Overall, the trend indicates that Ni will segregate towards subsurface sites to relax the local surface tension.

|  | CuNi(100) | | CuNi(110) | | CuNi(111) | |
|---|---|---|---|---|---|---|
| **Layer #** | DFT (J/m$^2$) | EAM (J/m$^2$) | DFT (J/m$^2$) | EAM (J/m$^2$) | DFT (J/m$^2$) | EAM (J/m$^2$) |
| **1st Layer** | 1.387 | 1.371 | 1.457 | 1.512 | 1.165 | 1.264 |
| **2nd Layer** | 1.360 | 1.365 | 1.430 | 1.505 | 1.150 | 1.256 |
| **3rd Layer** | 1.364 | 1.362 | 1.429 | 1.501 | 1.152 | 1.257 |

**Table A2:** Surface energies of Cu with 1x Ni dopant atom included on the surface, subsurface, or interior (1st, 2nd, or 3rd layer) to illustrate segregation trends from DFT and EAM

The monovacancy formation energies for pure Cu and pure Ni are 1.29 eV and 1.57 eV, respectively.

**Appendix B: Composition Profile Estimation MC Data**

Each step of the simulated annealing to equilibrium is dependent on the previous configuration (serial correlation), so the observed samples for calculating the estimated mean composition have a biased standard error that is overestimated by a naïve error calculation. To correct this, autocorrelation of subsequent frames in the MC process was estimated by assuming samples are generated by a first order autoregressive process (AR(1)):

$$X_t = c + \rho X_{t-1} + \varepsilon_{unbiased}$$

where $X_t$ is the estimated value at time t by the AR(1) model, c is an optional shifting constant (typically zero), and $\varepsilon$ the corrected error we wish to estimate. The process has an analytically known correction to find the unbiased, standard error of the estimated mean from n measurements and depends on $\rho$ to calculate the measurement error for each layer[55]:

$$\sigma_{unbiased} = \sigma_{biased} \sqrt{\frac{1 + 2\delta/n}{1 - \frac{2\delta}{n(n-1)}}}$$

$$\delta = \frac{(n-1)\rho - n\rho^2 + \rho^{n+1}}{(1-\rho)^2}$$

A custom Perl script was used to analyze the output data from MC, finding the composition in each layer and tracking it over all the annealed samples. These measurements were exported to a Python script which estimated values of $\rho$ for the data using the 'statsmodels' module, and the above equations estimated accurate errors for plots of the equilibrium composition profiles.

**Appendix C: Cluster Expansion Model for Kinetic Monte Carlo (KMC) simulation**

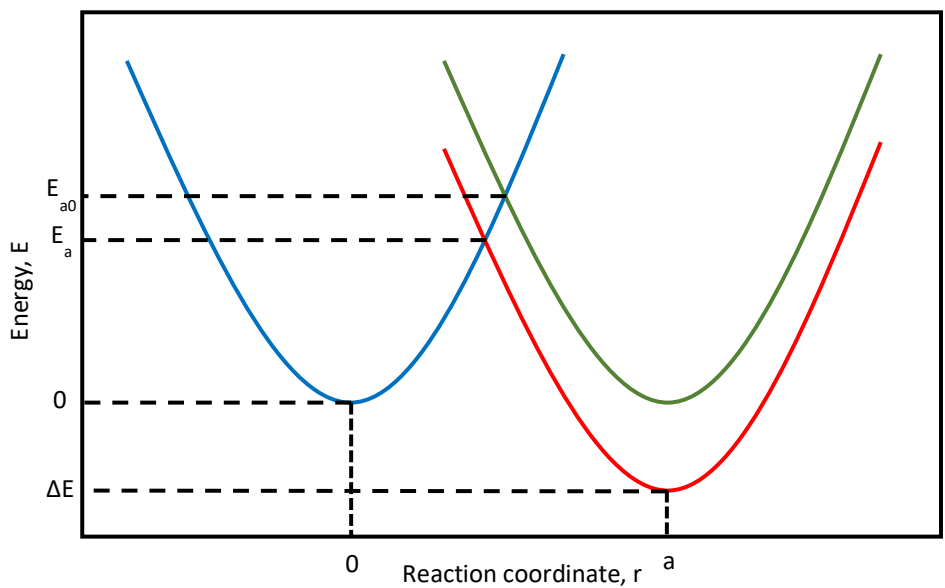

**Figure C1:** Parabolic construction showing energy diagram of a system. The blue curve is the initial state, the green curve is the final state with the same energy as the blue curve while the red curve is the final state with lower energy than the blue curve (Recreation)[56].

The reaction coordinate between two states is identified and analyzed using high-temperature sampling and climbing image nudged elastic band (CI-NEB) in AKMC. Contrastingly, our KMC model approximates these transition energies using an equation parametrized from barriers already known from AKMC: in Figure C1, the energy of the red curve is lower than the green curve due to a change in local environment of the vacancy. As a result, the barrier is lower and the rate can be calculated analytically as [56]

$$E_a = \frac{(4E_{a0} + \Delta E)^2}{16E_{a0}}$$

$$\text{rate} = Ae^{\frac{-E_a}{kT}}$$

where $E_a$ is the barrier from one state to a lower-energy state (blue and red), $E_{a0}$ is the barrier between two isoenergetic states (blue and green) also known as the intrinsic barrier, $\Delta E$ is the energy difference between the initial and the final state, A is the pre-factor in $s^{-1}$, k is the Boltzmann constant 8.617 $eV.K^{-1}$, and T is temperature in K. The intrinsic barrier $E_{a0}$ is obtained from AKMC data in Table C1.

**Table C1:** Intrinsic barrier of a vacancy migration event taken from AKMC data

| Location of the vacancy | Surface | | Subsurface | |
|---|---|---|---|---|
| Moving atom | Cu | Ni | Cu | Ni |
| Intrinsic barrier $E_{a0}$ (eV) | 0.453 | 0.487 | 0.635 | 0.681 |

To acquire $\Delta E$, the energy of each state is needed for the equation

$$\Delta E = E_{v,i} - E_{v,j}$$

where $E_{v,i}$ is the total energy of the configuration with the vacancy at location i (final) and $E_{v,i}$ is the total energy of the configuration with the vacancy at location j (initial). Vacancy formation energy is calculated by

$$E_f = E_v - E_0 + \mu_a$$

where $E_f$ is the vacancy formation energy, $E_v$ is the total energy of the configuration with the vacancy, $E_0$ is the total energy of the configuration without the vacancy (initial configuration) and $\mu_a$ is the chemical potential of the atom removed from the initial configuration to make the vacancy.

The vacancy formation energy can then be used to determine ΔE:

$$\Delta E = (E_{f,i} + E_0 - \mu_a) - (E_{f,j} + E_0 - \mu_a) = E_{f,i} - E_{f,j}$$

Vacancy formation energy is predicted by the cluster expansion method, which samples multiple alloy structures uniformly at random vacancies with different local environments. The energies of these structures were calculated and used to parametrize the model shown in Figure C2.

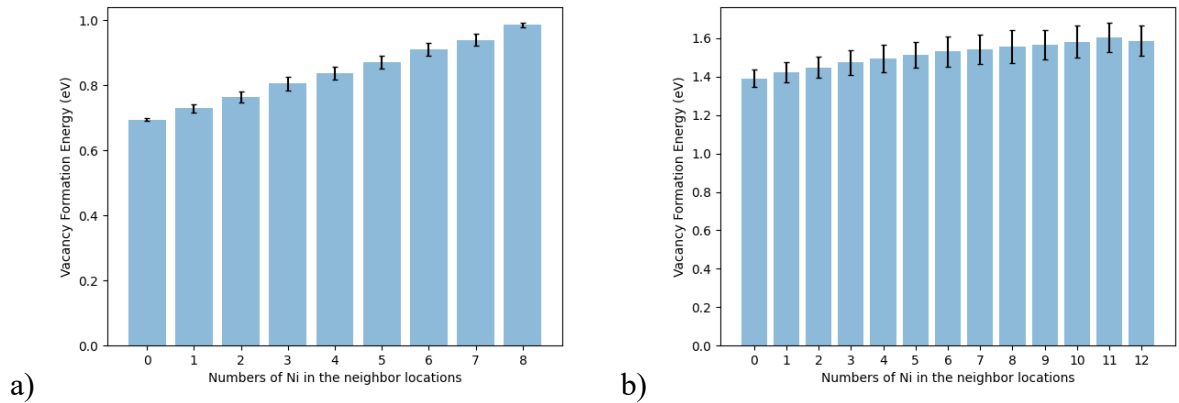

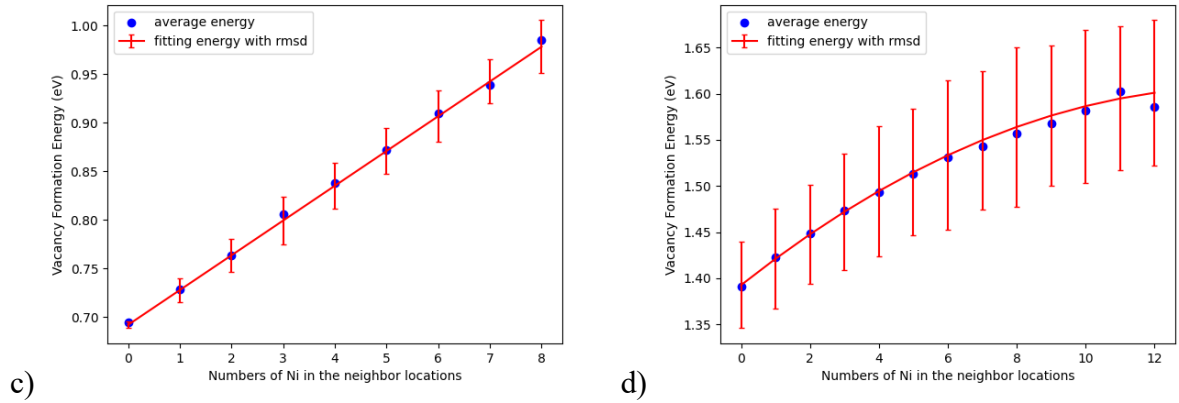

c)   d)

**Figure C2:** Average and standard deviation values of vacancy formation energy at (a) surface and (b) subsurface. Cluster expansion fitting model for the vacancy formation energy at (c) surface and (d) subsurface.

While the cluster expansion model for KMC works for surface-surface and subsurface-subsurface diffusion, it fails to illustrate correct surface-subsurface migrations (Figure C2b) since the identity of the migrating atom (counter to the vacancy movement) is not included in this model: Ni requires higher barriers (~0.95-1.0 eV) than Cu (~0.8-0.9 eV) to move from surface to subsurface according to AKMC data. To fix this significant qualitative limitation of the cluster expansion model, a new definition for the barriers of a vacancy migration from surface to subsurface is proposed based on the AKMC data:

$$E_{a,surf-sub} = \Delta E + E_{a,sub-surf}$$

The AKMC data shows that the barriers of a vacancy migration from surface to subsurface are very similar to the energy difference of final and initial states ($\Delta E$ in the equation above), and values of , $E_{a,sub-surf}$ are very small (0.049 eV for moving Cu atom and 0.067 eV for moving Ni atom) regardless of local environment. Figure C3 depicts an interpretation to calculate the barriers of a vacancy migration from surface to subsurface by adding the difference in energy, ΔE, between two states to $E_{a,sub-surf}$.

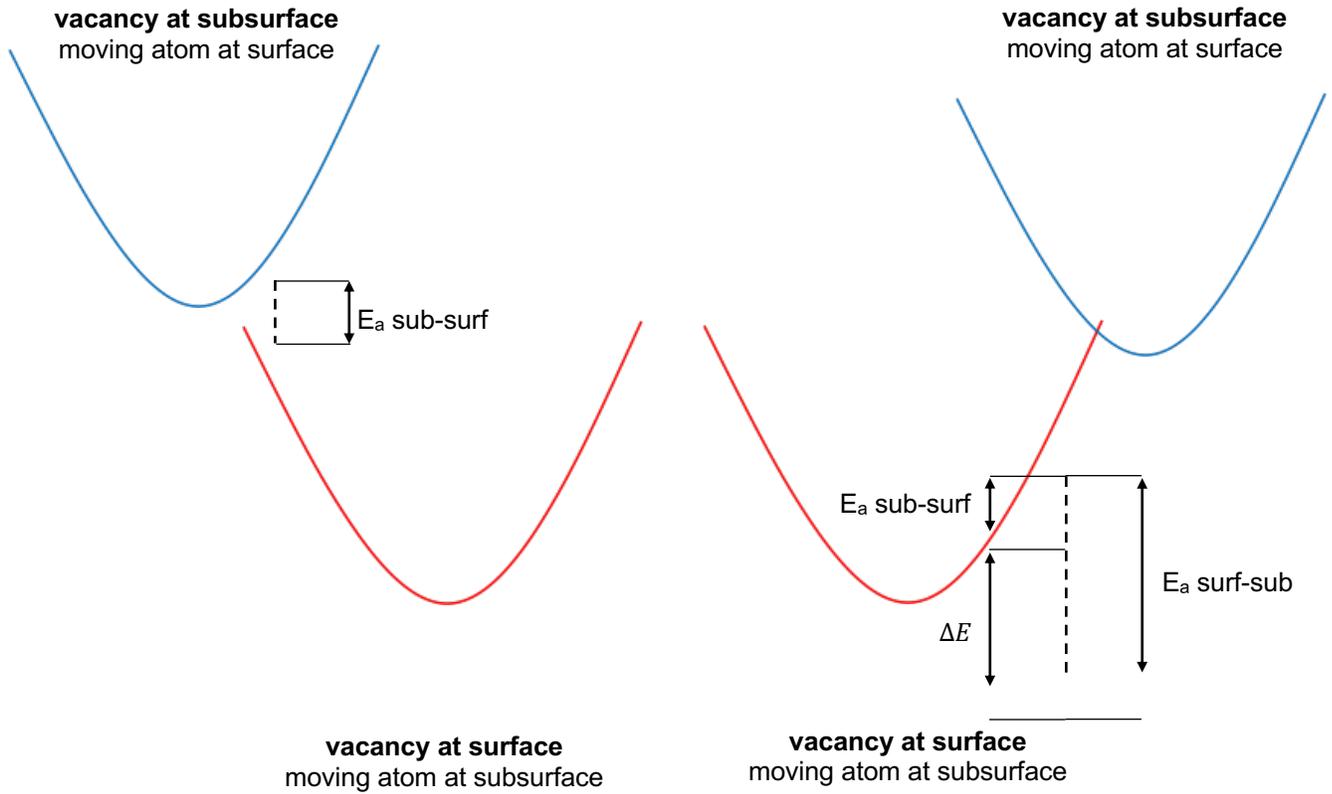

**Figure C3.** An illustration to determine the barrier of vacancy migration from surface to subsurface using the existed barrier of vacancy migration from subsurface to surface.

By the new definition, the barrier difference between Ni and Cu originates from the energy difference of two states ΔE, and its correction is significant to obtain correct dynamics. To include this information, ΔE was calculated from EON, where AKMC runs, then compared with the numerically calculated ΔE from cluster expansion. This discrepancy is illustrated in Figure C4[50]: the errors are approximately 0.09 eV for Cu and 0.24 eV for Ni, so these correction values were chosen to improve the cluster expansion model.

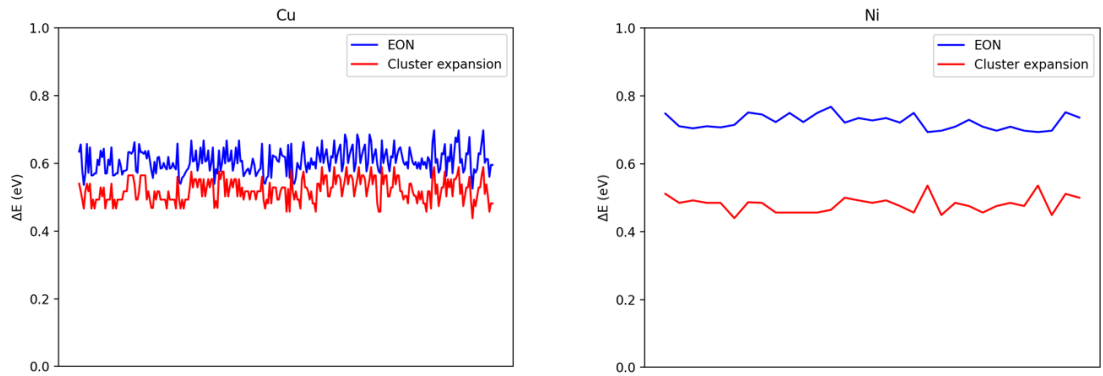

**Figure C4:** An illustration of discrepancy between ΔE from EON and from cluster expansion

**Appendix D: Comparison of AKMC and KMC results**

Figure D1 presents the energy profiles obtained from AKMC and KMC simulations that reach timescale of 300 μs, realizing 144,597 and 124,650 transition states respectively. Both methods show two Ni segregation events away from the surface at 100 ~ 125 μs and 170 ~ 200 μs, which corresponds to the energy drops in Figure. This comparison supports that our KMC model can simulate the similar behaviors that were observed in AKMC.

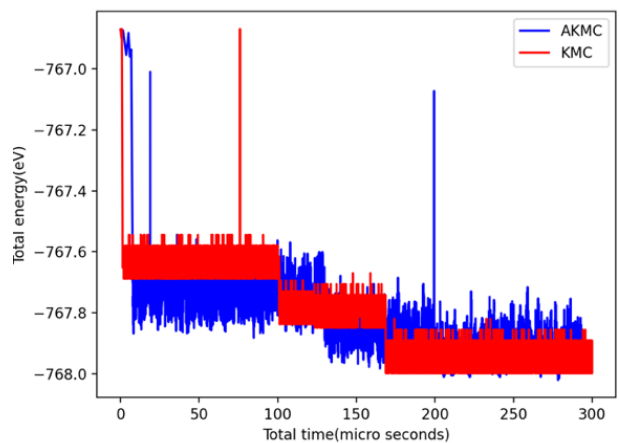

**Figure D1:** Energetic profiles of AKMC (blue) and KMC (red) simulations with respect to the total time of 300 μs.